\documentclass[12pt]{article}

\usepackage{graphicx}
\usepackage{color}
\usepackage{amsfonts}
\usepackage{amssymb}
\usepackage{subfigure}
\usepackage{hyperref}

\title{On low-energy predictions of unification models \\ inspired by F-theory}
\date{}
\author{T. Jeli\'{n}ski,  J. Pawe{\l}czyk and K. Turzy\'{n}ski\\{\small {}}\\{\small {\it Institute of Theoretical Physics, Faculty of Physics, University of Warsaw,}}\\{\small {\it ul.\ Ho\.za 69, 00-681 Warsaw, Poland}}\\}

\def\beq{\begin{equation}}
\def\eeq{\end{equation}}

\def\beqa{\begin{eqnarray}}
\def\eeqa{\end{eqnarray}}

\def\bar{\begin{array}}
\def\ear{\end{array}}

\def\ben{\begin{enumerate}}
\def\een{\end{enumerate}}

\def\bit{\begin{itemize}}
\def\eit{\end{itemize}}

\def\ZZ{\mathbb{Z}}

\def\a{\alpha}
\def\b{\beta}
\def\g{\gamma}

\def\D{\Delta}

\def\la{\lambda}
\def\La{\Lambda}

\def\zdef{\mkern-1mu\,\hbox{$\,\raise.3pt\hbox{:}\mkern-5mu =\mkern1mu\,$}}

\def\pd{\partial}

\newcommand{\eqref}[1]{(\ref{#1})}

\def\nn{\nonumber}

\def\bino{\widetilde{\chi}^0_1}

\graphicspath{{plots/}}

\newcommand\ov[1]{{\overline #1}}
\newcommand\vev[1]{\ensuremath{\langle #1\rangle} {}}
\newcommand\lgut{\ensuremath{\La_{GUT}} {}}

\definecolor{orange}{rgb}{1,0.5,0}

\begin{document}

\maketitle

\abstract{\normalsize The aim of this paper is to discuss phenomenological consequences of
a particular unification model ($\mathbb{Z}_3$ model) inspired by F-theory. 
The most distinctive feature of this model is 
a variety of (cosmologically feasible) options for the NLSP and NNLSP, 
beyond the usually considered benchmark scenarios. 
}

\section{Introduction}

\normalsize With the LHC experiments laying siege to the Minimal Supersymmetric Standard Model (MSSM)
and excluding large portions of its parameter space (see e.g.\ \cite{Atlas-11-086,Sekmen:2011cz})
it becomes increasingly important to identify the patterns of the masses of the superpartners
motivated by fundamental theories and to confront them with the existing experimental data.
Some novel patterns of this kind can emerge in unification models inspired by
F-theory \cite{F-unify1,F-unify2,F-unify3,F-unify4}. The presented local models of \cite{Heckman:2009mn} have several distinct features with possibly interesting impact on the low energy physics: extra $U(1)$ gauge symmetries which constrain the possible couplings and may lead to suppression of baryon number violating processes \cite{Pawelczyk:2010xh}, rich messenger sector and several scalars whose interactions can lead to supersymmetry (SUSY) breaking; some of these scalars may also be the dark matter particles. These models employ gauge mediation of supersymmetry
breaking (see e.g.\ \cite{Giudice:1998bp} for a review) to generate the soft supersymmetry
breaking masses for the scalar particles and the gauginos.
Preliminary studies of phenomenology of these models have been done in \cite{Heckman:2010xz} yielding bino-like neutralino
or  a stau as the next-to-lightest supersymmetric particle (NLSP), which is very similar to the usual models of gauge mediated supersymmetry breaking (GMSB); as always in these models the lightest supersymmetric particle is a GeV-scale gravitino. More recently,
the influence of some exotic matter on GMSB has been analyzed in \cite{Dolan:2011aq}.

In this letter we revisit the issue of the lowest mass SUSY  particles in the $\ZZ_3$  model proposed in  \cite{Heckman:2009mn,Pawelczyk:2010xh}. 
Analyzing all couplings allowed by the symmetries of the model, we find
that certain Yukawa-type couplings between the messengers and the matter fields
result in new contributions to the soft masses and $A$-terms for the scalar particles. 
Thanks to these corrections, new patterns of the superpartner masses become possible,
including the cases of a relatively light stops, sneutrinos and selectrons/smuons which has not shown up too often in previous studies.

This letter is organized as follows. After a brief description of the model in Section \ref{sec:model},
we present the expressions for the soft supersymmetry breaking terms at the messenger scale in Section \ref{sec:soft}. We study phenomenological consequences of the model by performing a random scan over its parameters; the scan is described in Section \ref{sec:num}. Among the results we find many
new patterns of the mass spectra of the superpartners and discuss their viability in Section \ref{sec:disc}. We conclude in Section \ref{sec:conc}.
 
 
\section{The model}
\label{sec:model}

The details of the construction of the Dirac $\mathbb{Z}_3$ model can be found in \cite{Heckman:2009mn,Pawelczyk:2010xh}, but for convenience we shall present a synopsis here. 
The model contains two extra local $U(1)$ symmetry groups,
whose gauge bosons acquire masses of the order of the GUT scale (denoted here by $\lgut$) in Green-Schwarz mechanism. Thus at low energies the $U(1)$'s provide extra selection rules for the construction of Lagrangian.
With the exception of the Higgs fields, all matter fields are arranged in complete $5$, $\overline{5}$, $10$ representations of $SU(5)$.  The Higgses $H_u,\ H_d$ are just the MSSM doublets
and belong to incomplete $5$ and $\overline{5}$ representations, and we shall identify $H_u=(5_H)_2,\ H_d=(\ov5_H)_2$.
The matter content is summarized in Table \ref{t:maco}. 
\begin{table}[h!]
\renewcommand{\arraystretch}{1.2}
\beqa
\hspace{6mm}\begin{tabular}
[c]{|c|c|c|c|c|c|c|c|}
\hline
&\multicolumn{4}{|c|}{MSSM}\\
\cline{2-5}
& $10_{M}$ & $\overline{5}_{M}$& $5_{H}$ & $\overline{5}_{H}$\\\hline
$U(1)_{PQ}$ & $+1$ & $+1$  & $-2$ & $-2$ \\\hline
$U(1)_{\chi}$ & $-1$  & $+3$ & $+2$ & $-2$\\\hline
\end{tabular}
\renewcommand{\arraystretch}{1.194}
\begin{tabular}[c]{|c|c|c|c|c|c|c|c|}\hline
\multicolumn{8}{|c|} {Exotic}\\\hline
$Y_{10}$ & $Y_{10}'$  &$Y_{\overline{10}},\ Y_{\overline{10}}'$ &  $Y_{\overline{5}}$ &$Y_{5}$  &$X$ & $N_R$ &$\overline D_1$ \\ \hline
$+1$ &$0$ &$+3$ & $+1$ & $+3$ & $-4$ & $-3$ &$-1$ \\\hline
$-1$  &$+4$&$+1$  & $+3$ & $-3$& $0$ & $-5$ &$+5$ \\\hline
\end{tabular}
\nn
\eeqa
\hspace{10mm}\parbox{16.5cm}{\caption{Matter content of the model. The columns marked as `MSSM'  show the $U(1)_{PQ}$ and $U(1)_{\chi}$ charges of the $SU(5)$ multiplets containing the MSSM fields. The charges of the remaining fields are shown under the label `Exotic'.}\label{t:maco}}
\end{table}

We must recall that in F-theory case the effective Lagrangian contains all the invariant couplings, including the Yukawa interactions in the superpotential and the trilinear terms in K\"ahler potential (divided by the GUT scale $\La_{GUT}$). For simplicity we shall take the respective couplings to be of the order one.
The messenger fields are $Y_5,\ Y_{10},\ Y_{10}'$, where the subscript indicates the $SU(5)$ representation.
Their masses are generated via vacuum expectation values of spurion superfields $X,\ N_R$ coupled as\footnote{We have assumed that coupling of $Y_{\overline{10}}, Y_{\overline{10}}'$ is diagonal. This  simplifies the model and precludes large 1-loop contributions to soft masses from $U(1)_Y$ $D$-terms \cite{Dimopoulos:1996ig}.}:
\beqa\label{WS}
W_S=X(Y_5Y_{\overline{5}}+Y_{10}Y_{\overline{10}})+N_RY'_{10}Y'_{\overline{10}} \, .
\eeqa
The 
superpotential describing interactions between the messengers and the matter fields
reads:
\beqa
W_Y=\frac{\lambda_1}2 (5_H)_2Y_{10}Y_{10}+\lambda_2 (\ov5_H)_2Y_{\ov5} Y_{10}+
\lambda_3 (5_H)_2{10}_MY_{10}+\lambda_4(\ov5_H)_2Y_{\ov5}{10}_M+
\lambda_5(\ov5_H)_2Y_{10}{\ov5}_M\,.
\label{Wh}
\eeqa
Note that the messenger $Y_{10}'$ does not show up in $W_Y$, hence  its only couplings of interest are those with vector supermultiplets. However, we shall argue below that $F_{N_R}$ is nearly zero, so there are no
mass splittings between the bosonic and fermionic components of $Y_{10}'$ and this multiplet
can be ignored in the analysis of supersymmetry breaking.

Finally the hidden sector consists of spurions ${X,N_R,\overline{D}_1}$ which are singlets of $SU(5)$ and whose nonzero $F$-terms break SUSY. Nonvanishing $F_{\overline{D}_1}$ generates B/L violating processes \cite{Pawelczyk:2010xh} while nonvanishing $F_{N_R}$ triggers breaking of the Standard Model gauge group by sleptons due to K\"ahler potential term \cite{Brahm:1990xx}
\beq
K \supset
\frac{1}{\lgut}H_d^\dagger L N_R+\mathrm{h.c.}\,,
\eeq
thus we set them to zero.
It follows that in the present analysis we can ignore $N_R,\overline{D}_1$ and   $Y'_{10}$ messengers and $X$ is the only  relevant spurion.

The model has three characteristic mass scales above the weak scale: 
$\lgut$, $M_Y=\vev X$ and $\xi\equiv F_X/\vev X$. 
The first one is the unification scale, $\lgut\approx 2\times 10^{16}\,\textrm{GeV}$, which we also
identify with the mass $M_{PQ}$ of the heavy gauge bosons of the extra $U(1)$'s.  
$M_Y$ is  the messengers scale and we
assume it in the range
$10^{13}\,\textrm{GeV}\lesssim M_Y\lesssim10^{14}\,\textrm{GeV}$.
The last one, $\xi$,  is related to SUSY breaking scale $\sqrt{F_X}$. We shall not discuss the dynamics which triggers SUSY breaking (the simplest choice is the Polonyi model with linear SUSY breaking term induced e.g. by D-brane instantons)
but assume for simplicity that SUSY is broken by $F$-term of the spurion $X$.  The value
$\sqrt{F_X}=10^{9.5}\,\textrm{GeV}$ was fixed for the gravitino mass $m_{\widetilde{G}}\sim F_X/M_P$ to be of the order 1 GeV \cite{Ibe:2007km}.


\section{Soft terms at the messenger scale}
\label{sec:soft}

Supersymmetry is broken in the hidden sector by the $F$-term of the spurion $ X $.
In general, SUSY breaking can be transmitted to the visible sector by a variety
of mechanisms. 
Due to couplings of messengers and matter in superpotential, 
our model represents a full-fledged mechanism of so-called deflected gauge mediation \cite{Chacko:2001km}. 
Gaugino masses arise from 1-loop diagrams, scalar masses mainly come
from 2-loop diagrams and there are 1-loop contributions to $A$-terms
originating from direct couplings between the messengers and the matter fields
\cite{Chacko:2001km,Giudice:1997ni,Shadmi:2011hs,Evans:2011be}.
Below we summarize these results at the messenger scale $M_Y$,
adopting a widely used convention $\alpha_g=g^2/4\pi$, where $g$
is a coupling constant.



Gaugino masses are given by standard 1-loop expressions:
\beq
M^{(r)}_{\la}=\frac{\a_r}{4\pi}n_X\xi \, ,
\eeq
where $r=1,2,3$ corresponds to the gauge group $U(1)_Y$, $SU(2)_L$ and $SU(3)_C$ of the
Standard Model (we use the GUT normalization for the hypercharge) and $n_X=4$
is twice the sum of the Dynkin indices of the messenger fields coupling to spurion $X$.



Soft SUSY breaking mass terms of scalars $\phi\in\{H_u,H_d,L,E,Q,U,D\}$ can be written as
\beq
\label{eq:mphi_1}
m_\phi^2=m_{\phi,g}^2+m_{\phi,\lambda}^2 + m_{\phi,PQ}^2 \, .
\eeq
Here $m^2_{\phi,g}$ are standard 2-loop gauge mediation results  induced by gauge interactions transmitting SUSY breaking from messenger sector
\beqa
m_{\phi,g}^2&=&2\sum\limits_{r=1}^3C_2^r(\phi)\frac{\a_r^2}{(4\pi)^2}n_X\xi^2,
\eeqa
where $C_2^r(\phi)$ are quadratic Casimir operators of the representation of $\phi$ under $r$-th gauge group.
The contributions
$m^2_{\phi,\lambda}$ in (\ref{eq:mphi_1}) 
are 2-loop terms induced by the Yukawa-type couplings in (\ref{Wh})
\cite{Chacko:2001km}:
\beqa
m_{\phi,\lambda}^2=-\frac{1}{4}\sum\limits_{\lambda}\left(\frac{\pd\D\g_\phi}{\pd \lambda}\b^{+}_{\lambda}+\mathrm{h.c.}
\right)\xi^2 \, ,
\label{eq:mphih}
\eeqa
where $\lambda$ denote all the Yukawa couplings in the superpotential.
In (\ref{eq:mphih}) $\b^+$ are the beta functions above the messenger mass scale $M_Y$, $\D\g_\phi=\g^+_\phi-\g^-_\phi$,  $\g^+_\phi$ is the anomalous dimension of $\phi$ above scale $M_Y$ and  $\g^-_\phi$ is the anomalous dimension of $\phi$ below scale $M_{Y}$. Discontinuities of $\g_\phi$ are due to the absence of the messengers in the effective action below scale $M_Y$. 
Note that 1-loop contributions to the soft SUSY breaking scalar masses (\cite{Dine:1996xk}) 
are negligibly small compared to $m^2_{\phi,g}$ and can be safely ignored.
We have checked that 
all $\lambda_i$ couplings can significantly change the pattern of superpartner masses. 
With the recipe outlined above, eq.\ (\ref{eq:mphih}) can be rewritten in the following
form:
\beq
\label{eq:mphih2}
m_{\phi,\lambda}^2 = \frac{\xi^2}{16\pi^2}\sum_{A}^{} \kappa_{\phi,A} f_A(\alpha_i) \, .
\eeq
In this formula, the functions $f_A$ are 
appropriate products of
$\alpha_{\lambda_i}$ with $i=1,\ldots,5$, $\alpha_{y_j}$ with $j=u,d,e$ 
(or square roots thereof)
and three combinations of the
gauge couplings which we denote by
$\alpha_G'=(13/60)\alpha_1+(3/4)\alpha_2+(4/3)\alpha_3$, $\alpha_G''=(7/60)\alpha_1+(3/4)\alpha_2+(4/3)\alpha_3$ and $\alpha_G'''=(1/20)\alpha_1+(1/12)\alpha_2$. 
These functions and
the values of the numerical coefficients $\kappa_{\phi,A}$ in the expansion
(\ref{eq:mphih2}) are given in Table \ref{t1}.

%


The contribution $m_{\phi,PQ}^2$ in (\ref{eq:mphi_1}) comes from an
operator generated at the tree level by exchange of the heavy gauge boson of anomalous $U(1)_{PQ}$ symmetry:
\beqa\label{PQ}
\mathcal{L}\supset-\frac{g_{PQ}^2q_Xq_\phi}{M^2_{PQ}}\int d^4\theta X^\dag X \phi^\dag\phi.
\eeqa
It is known that this operator
can significantly affect the low-energy spectrum of the model 
\cite{Heckman:2010xz}.
We shall parametrize soft masses $m^2_{\phi,PQ}$ generated by \eqref{PQ} as
\beq
\label{eq:mphi_pq}
m^2_{\phi,PQ}= q_Xq_\phi\widetilde{\Delta}^2 \, ,
\eeq
where $q_X$, $q_{\phi}$ are $U(1)_{PQ}$ charges and $\widetilde{\Delta}=g_{PQ}\frac{|F_X|}{M_{PQ}}=g_{PQ}\frac{|F_X|}{\lgut}$. With given $F_X$ and $\lgut$ this quantity depends only on $g_{PQ}$, which we expect to be of the same order as the unified value of all gauge couplings at $\lgut$.
Thus the  natural scale for $\widetilde\Delta$ is
$\mathcal{O}(10^2\,\mathrm{GeV})$. 

Finally, there are $A$-terms generated by 1-loop diagrams involving
the Yukawa-type couplings in (\ref{Wh}). 
We can write their contributions to the trilinear terms in
the potential as:
\beqa\label{VA}
V&\supset& -\frac{y_u}{4\pi}(3\a_{\lambda_1}+9\a_{\lambda_3}+\a_{\lambda_4})\xi H_uQU-\frac{y_d}{4\pi}(4\a_{\lambda_2}+\a_{\lambda_3}+5\a_{\lambda_4}+6\a_{\lambda_5})\xi H_dQD\nonumber\\
&&-\frac{y_e}{4\pi}(4\a_{\lambda_2}+6\a_{\lambda_4}+5\a_{\lambda_5})\xi H_dLE.
\eeqa
These results specify the initial conditions which we shall use in the following
section to study the phenomenological aspects of the model.

\newpage

\begin{table}[h!]
\begin{center}
{
\begin{tabular}{|c|ccccccc|}
\hline
\phantom{$\downarrow$ $f_A$} $\phi\rightarrow$& $H_u$ & $H_d$ & $L$ & $E$ & $Q$ & $U$ & $D$ \\
$\downarrow$ $f_A$ \phantom{$\phi\rightarrow$} &&&&&&&\\ 
%
\hline
$\alpha_{\lambda_1}^2$ & 18 & 0 & 0 & 0 & 0 & 0& 0 
\\
$\alpha_{\lambda_1}$ $\alpha_{\lambda_2}$& 3 & 3 & 0 & 0 & 0 & 0& 0 
\\
$\alpha_{\lambda_1}$ $\alpha_{\lambda_3}$& 72 & 0 & 0 & 0 & 6 & 12& 0 
\\
$\alpha_{\lambda_1}$ $\alpha_{\lambda_5}$& 3 & 3 & 0 & 0 & 0 & 0& 2 
\\
$\alpha_{\lambda_2}^2$& 0 & 28 & 0 & 0 & 0 & 0& 0 
\\
$\alpha_{\lambda_2}$ $\alpha_{\lambda_3}$& 3 & 3 & 0 & 0 & 0 & 2 & 0 
\\
$\alpha_{\lambda_2}$ $\alpha_{\lambda_4}$& 0 & 56 & 0 & 14 & 7 & 0& 0 
\\
$\alpha_{\lambda_2}$ $\alpha_{\lambda_5}$& 0 & 56 & 7 & 0 & 0 & 0 & 14 
\\
$\alpha_{\lambda_3}^2$ & 54 & 0 & 0 & 0 & 9 & 18 & 0 
\\
$\alpha_{\lambda_3}$ $\alpha_{\lambda_4}$& 3 & 3 & 0 & 0 & 2 & 0& 0 
\\
$\alpha_{\lambda_3}$ $\alpha_{\lambda_5}$& 3 & 3 & 0 & 0 & 0 & 2 & 2 
\\
$\alpha_{\lambda_4}^2$ & 0 & 28 & 0 & 14 & 7 & 0 & 0 
\\
$\alpha_{\lambda_4}$ $\alpha_{\lambda_5}$& 0 & 32 & 4 & 8 & 4 & 0 & 8 
\\
$\alpha_{\lambda_5}^2$ & 0 & 28 & 7 & 0 & 0 & 0 & 14 
\\
$\alpha_{\lambda_1}$ $\a_{y_u}$ & 0 & 0 & 0 & 0 & $-3$ & $-6$ & 0 
\\
$\alpha_{\lambda_2}$ $\a_{y_d}$ & 0 & 0 & 0 & 0 & $-4$ & 0 & $-8$ 
\\
$\alpha_{\lambda_2}$ $\a_{y_e}$ & 0 & 0 & $-4$ & $-5$ & 0 & 0 & 0 
\\
$\alpha_{\lambda_3}$ $\a_{y_u}$ & 0 & 0 & 0 & 0 & $-6$ & $-12$ & 0 
\\
$\alpha_{\lambda_3}$ $\a_{y_d}$ & 3 & $-3$ & 0 & 0 & 0 & 0 & $-2$ 
\\
$\alpha_{\lambda_4}$ $\a_{y_u}$ & $-3$ & 3 & 0 & 0 & 0 & $-2$ & 0 
\\
$\alpha_{\lambda_4}$ $\a_{y_d}$ & 0 & 0 & 0 & 6 & $-1$ & 0 & $-14$ 
\\
$\alpha_{\lambda_4}$ $\a_{y_e}$ & 0 & 0 & $-7$ & $-6$ & 1 & 0 & 0 
\\
$\alpha_{\lambda_5}$ $\a_{y_d}$ & 0 & 0 & 3 & 0 & $-7$ & 0 & $-2$ 
\\
$\alpha_{\lambda_5}$ $\a_{y_e}$ & 0 & 0 & $-3$ & $-14$ & 0 & 0 & 2 
\\
$\alpha_G'$ $\alpha_{\lambda_1}$ & $-12$ & 0 & 0 & 0 & 0 & 0 & 0 
\\
$\alpha_G'$ $\alpha_{\lambda_3}$ & $-24$ & 0 & 0 & 0 & $-4$ & $-8$ & 0 
\\
$\alpha_G''$ $\alpha_{\lambda_2}$ & 0 & $-12$ & 0 & 0 & 0 & 0 & 0 
\\
$\alpha_G''$ $\alpha_{\lambda_4}$ & 0 & 0 & 0 & 0 & $-4$ & 0 & 0 
\\
$\alpha_G''$ $\alpha_{\lambda_5}$ & 0 & $-12$ & 0 & 0 & 0 & 0 & $-8$ 
\\
$\alpha_G'''$ $\alpha_{\lambda_2}$ & 0 & $-36$ & 0 & 0 & 0 & 0 & 0 
\\
$\alpha_G'''$ $\alpha_{\lambda_4}$ & 0 & $-36$ & 0 & $-72$ & 0 & 0 & 0 
\\
$\alpha_G'''$ $\alpha_{\lambda_5}$ & 0 & $-36$ & $-36$ & 0 & 0 & 0 & 0 
\\
$\sqrt{\a_{\lambda_1}\a_{\lambda_2}\a_{\lambda_3}\a_{\lambda_4}}$ & 6 & 6 & 0 & 0 & 2 & 0 & 0 
\\
$\sqrt{\a_{\lambda_1}\a_{\lambda_3}\a_{\lambda_3}\a_{y_u}}$ & 18 & 0 & 0 & 0 & 0 & 0 & 0 
\\
$\sqrt{\a_{\lambda_1}\a_{\lambda_3}\a_{\lambda_5}\a_{y_d}}$ & 6 & 0 & 0 & 0 & 0 & 0 & 0 
\\
$\sqrt{\a_{\lambda_2}\a_{\lambda_3}\a_{\lambda_4}\a_{y_u}}$ & 0 & 6 & 0 & 0 & 0 & 0 & 0 
\\
$\sqrt{\a_{\lambda_2}\a_{\lambda_4}\a_{\lambda_5}\a_{y_d}}$ & 0 & 18 & 0 & 0 & 0 & 0 & 0 
\\
$\sqrt{\a_{\lambda_2}\a_{\lambda_4}\a_{\lambda_5}\a_{y_e}}$ & 0 & 6 & 0 & 0 & 0 & 0 & 0 
\\
$\sqrt{\a_{\lambda_3}\a_{\lambda_5}\a_{y_u}\a_{y_d}}$ & 0 & 0 & 0 & 0 & $-2$ & 0 & 0 
\\
\hline
\end{tabular}
}
\end{center}
\caption{Numerical coefficients $\kappa_{\phi,A}$ in (\ref{eq:mphih2}). \label{t1}}
\end{table}

\section{Numerical analysis}
\label{sec:num}

Adopting the initial conditions for the soft SUSY breaking masses presented in Section \ref{sec:soft},
we now turn to studying the phenomenological consequences of such an ansatz.
We compute the low-energy spectrum and the electroweak symmetry breaking
with an appropriately modified {\tt SuSpect} code \cite{Djouadi:2002ze}. We work in the
approximation of vanishing Yukawa couplings of the first two generations of fermions, so
the MSSM mass spectra we obtain are degenerate for these generations. In the following,
we shall refer to sfermions of these generations with the name of the first generation.

In order to make the large parameter space manageable, we choose 
$\xi=10^5\,\mathrm{GeV}$ and $M_Y=10^{14}\,\mathrm{GeV}$. 
Such a range is consistent with
models of gravitational stabilization of the SUSY breaking vacuum put forward in 
\cite{Kitano:2006wz}, which have $\vev X\sim M_Y\sim \Lambda_{GUT}^2/M_P$. 
This choice of $\xi$ leads to the bino/wino/gluino masses
of approximately \mbox{0.55/1.0/2.7 TeV}.
We also take
$\mathrm{sgn}(\mu)=+1$. 

Parameters $\lambda_i$ 
are varied between 0.2 and 1.5 with a flat prior, but the maximal values of $\lambda_1$ and $\lambda_3$
consistent with the bounds discussed below are 1.1 and 0.9, respectively. 
The nonzero lower limit reflects the assumption that natural values of the couplings $\lambda_i$
lie close to unity, so these couplings should not be fine tuned to very small values.
The values of $\tan\beta$ are drawn from the range $5-45$ with a flat prior.
We also scan $\widetilde\Delta^2$ between $2.5\times 10^3\,\mathrm{GeV}^2$ and $2.5\times 10^4\,\mathrm{GeV}^2$ with a flat prior.

We impose a number of constraints on the obtained mass spectra. We require that the scalar potential
is bounded from below and that there are no low lying color or charge breaking minima. The latter
leads to quite involved conditions \cite{LeMouel:2001sf}, but we checked and then used its
simplified (and more practical computationally) version implemented in {\tt SuSpect} which
does not introduce large errors. All models with tachyons in the spectrum, 
the light Higgs boson mass below 114 GeV 
and
$\mathrm{BR}(b\to s\gamma)$ lying outside the 2$\sigma$ range
$(2.87-4.33)\times 10^{-4}$ were discarded. We keep only models
in which the squark and gluino masses lie within the allowed $95\%$ CL range determined for a
simplified setup in Ref.\ \cite{Atlas-11-086}. Note that these bounds are roughly consistent with
those obtained in the phenomenological MSSM with 19 independent parameters \cite{Sekmen:2011cz}.

\begin{figure}[!h]
\subfigure[]{\includegraphics[scale=0.725]{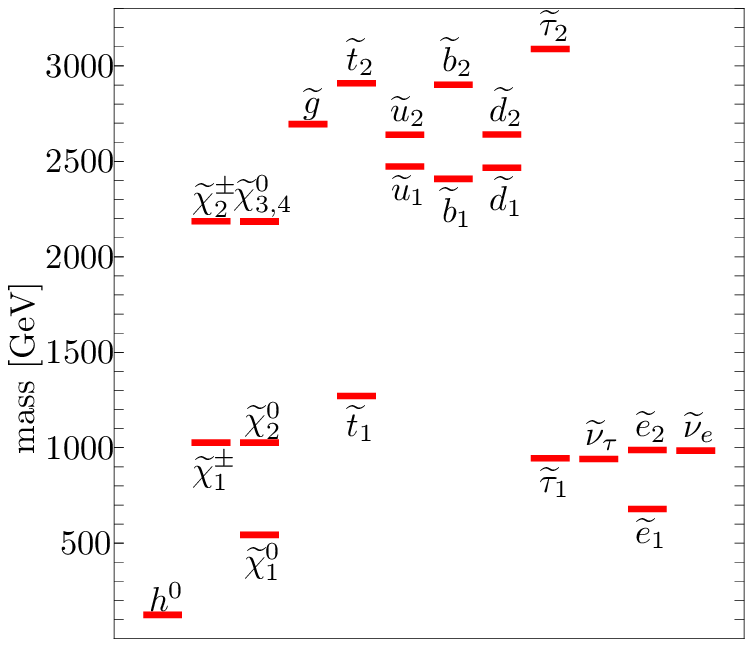}}
\subfigure[]{\includegraphics[scale=0.725]{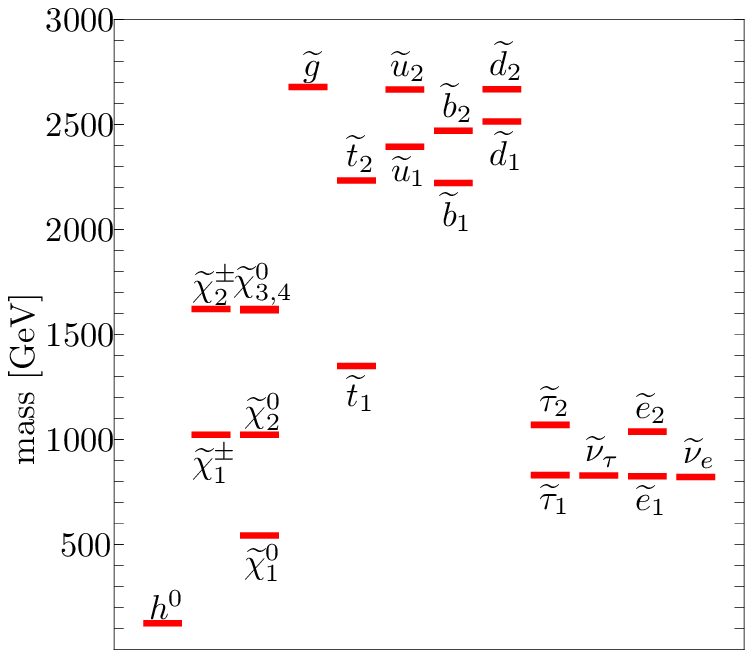}}
\subfigure[]{\includegraphics[scale=0.725]{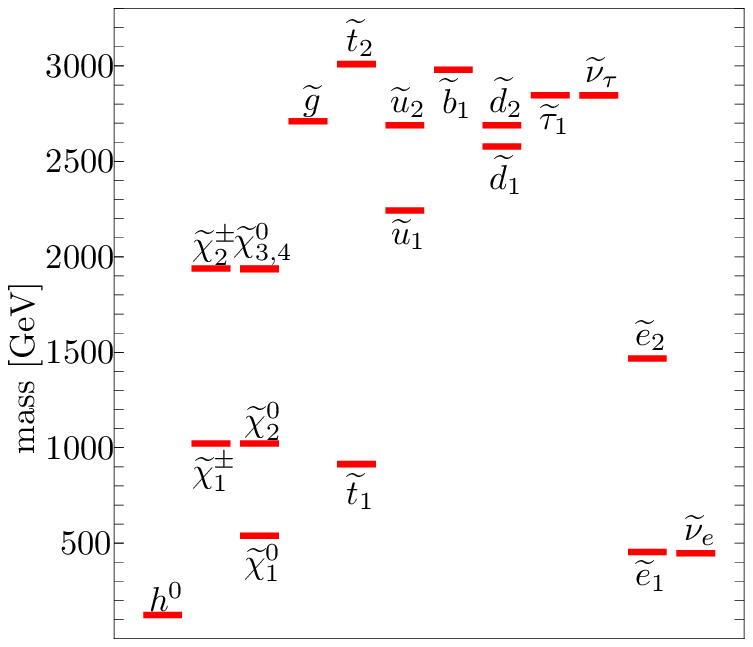}}
\subfigure[]{\includegraphics[scale=0.725]{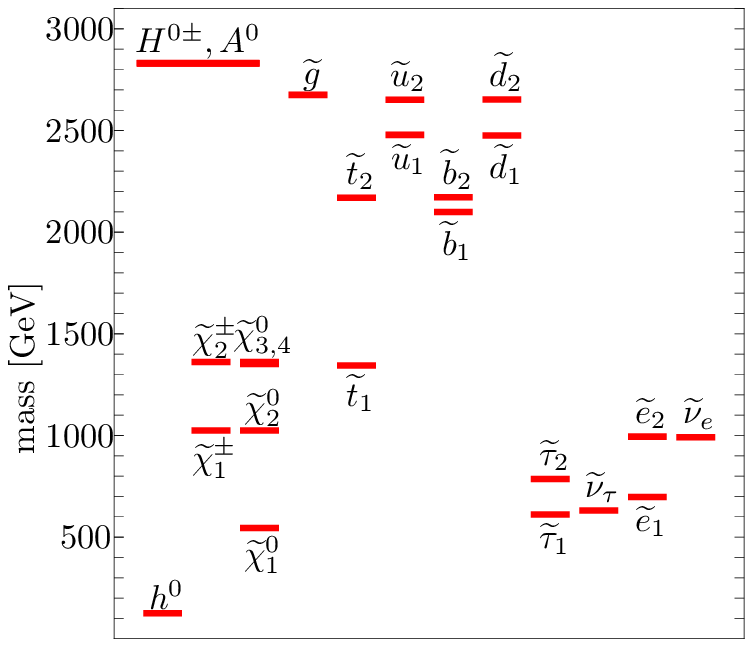}}
\subfigure[]{\includegraphics[scale=0.725]{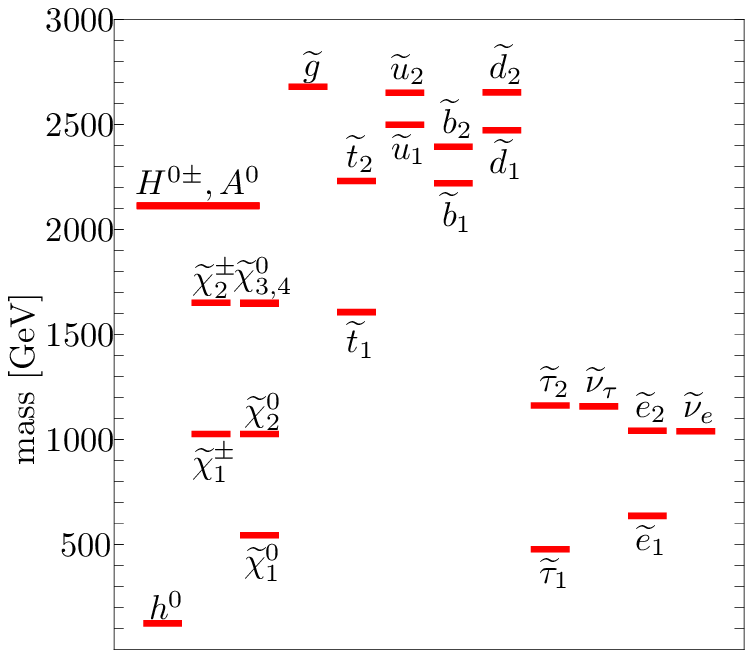}}
\subfigure[]{\includegraphics[scale=0.725]{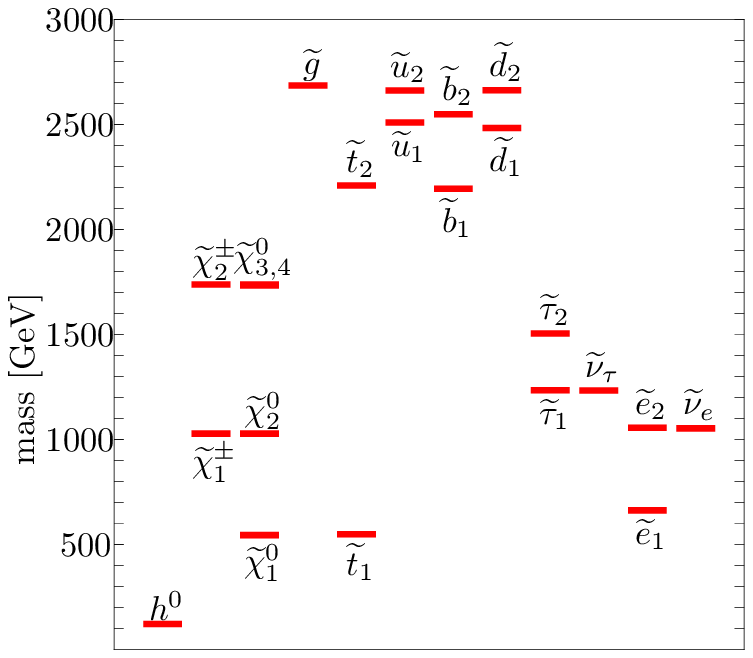}}
\subfigure[]{\includegraphics[scale=0.725]{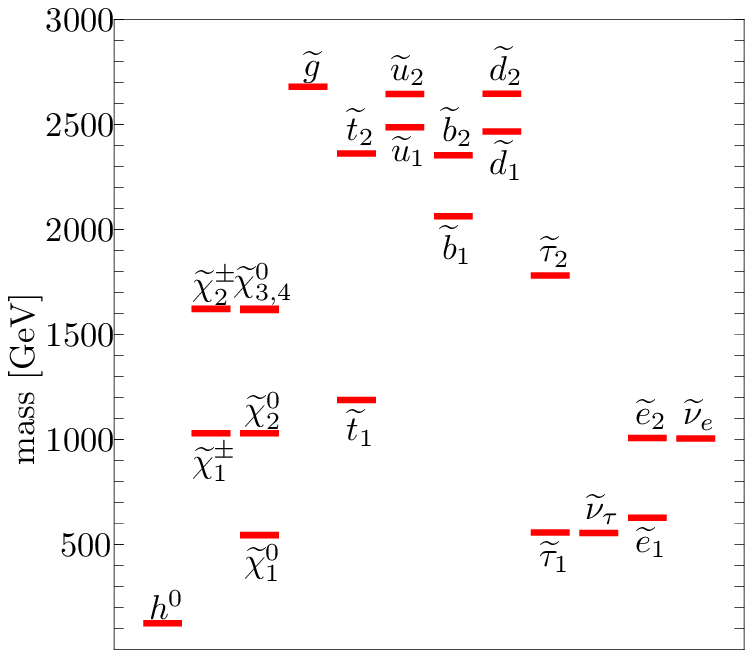}}
\subfigure[]{\includegraphics[scale=0.725]{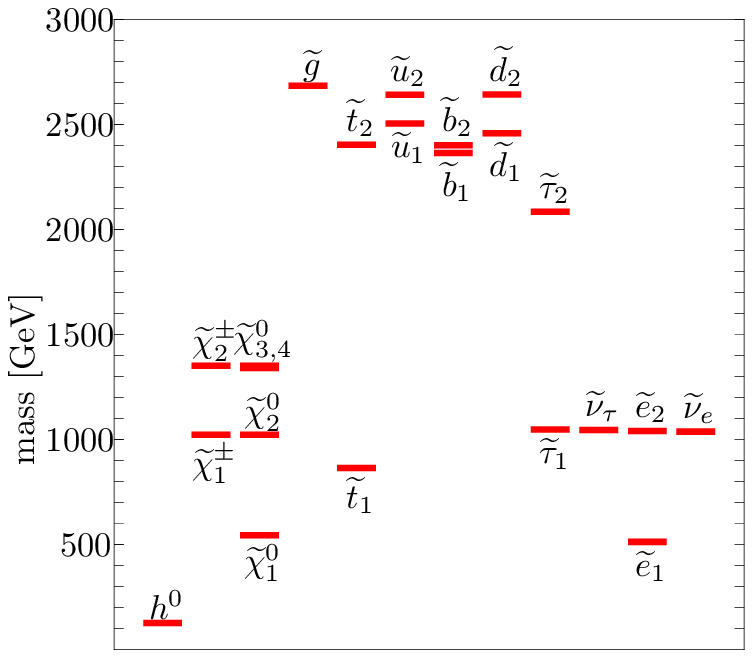}}
\subfigure[]{\includegraphics[scale=0.725]{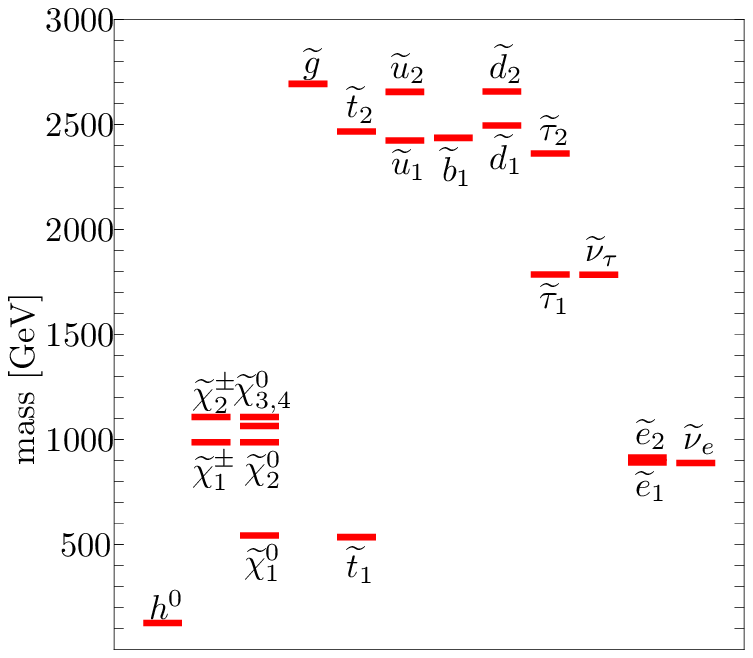}}
\subfigure[]{\includegraphics[scale=0.725]{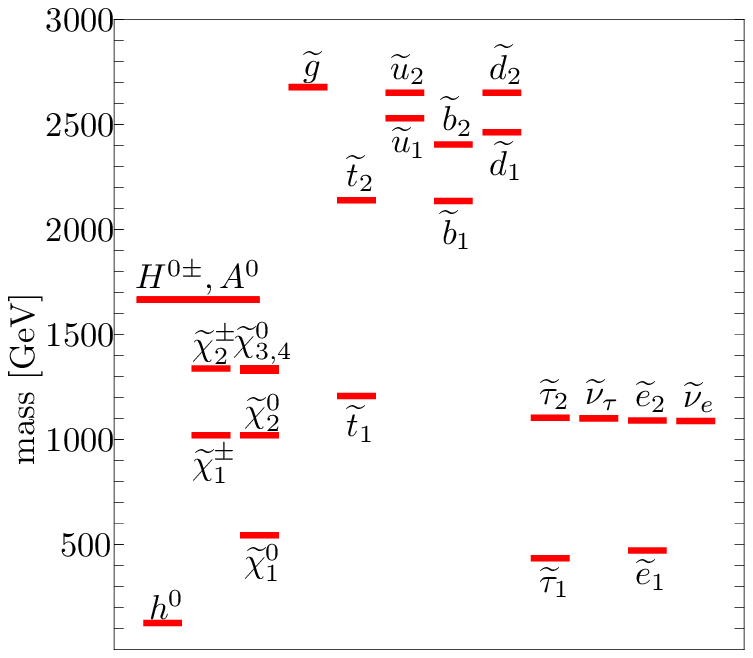}}
\hspace{8.9mm}\subfigure[]{\includegraphics[scale=0.725]{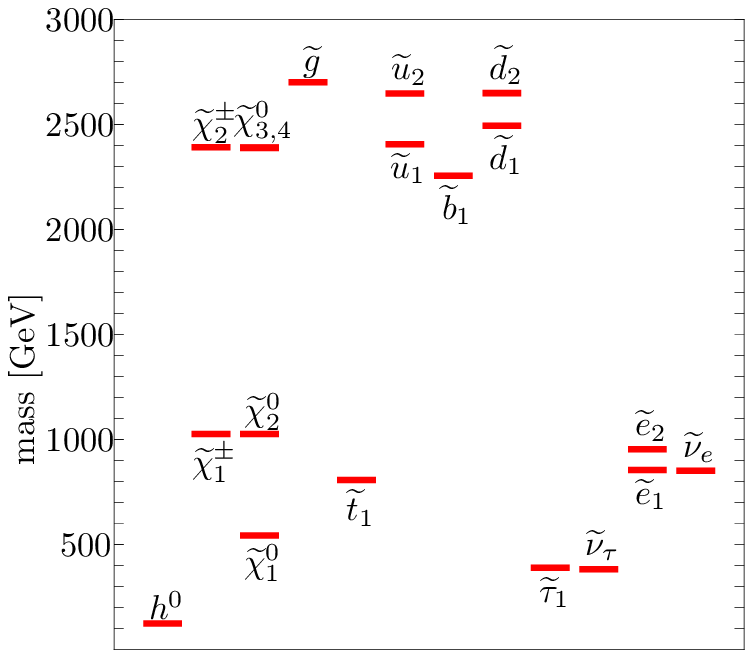}}
\hspace{8.9mm}\subfigure[]{\includegraphics[scale=0.725]{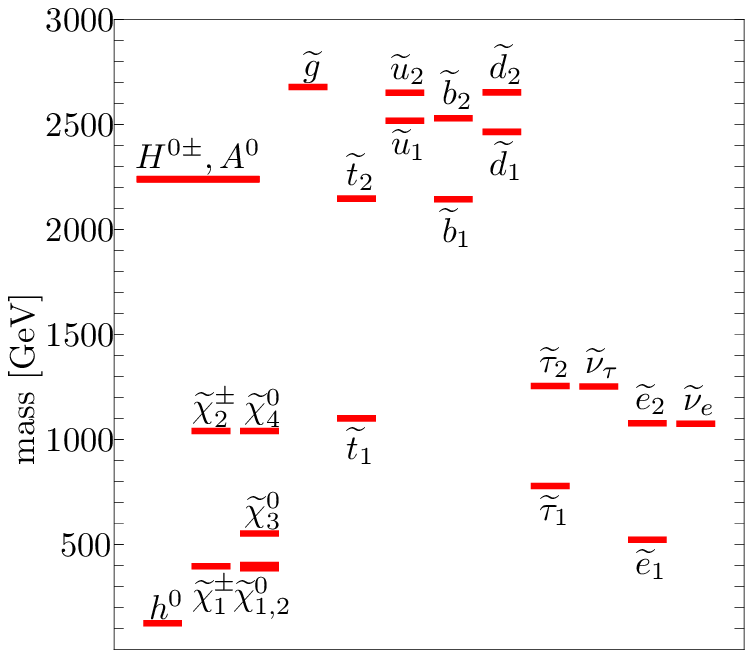}}
\phantom{a}\hspace{0.8cm}\parbox{16.5cm}{\caption{Examples of the mass spectra corresponding to parameters $\lambda_i$, $\widetilde{\Delta}$ and $\tan\beta$ given in  Table \ref{t:ex}.}\label{fig:ex}}
\end{figure}

The patterns of NLSP/NNLSP occurring in the scan with frequency bigger than 0.01 are shown in Table \ref{t:ex} and plotted in Fig. \ref{fig:ex}.
They encompass much wider variety of possibilities than usually considered
in models of GSMB. In particular, there are 3 classes of
novel patterns: selectron (N)NLSP, stop (N)NLSP and higgsino (N)NLSP. 
\begin{table}[!h]
\begin{center}
\begin{tabular}{|c|cc|ccccccc|cc|}
\hline
& {\footnotesize NLSP} & {\footnotesize NNLSP} & $\la_1$ & $\la_2$ & $\la_3$ & $\la_4$ & $\la_5$ 
& {\footnotesize $(\frac{10^3\widetilde{\D}}{\xi})^2$} & { $\tan\beta$} &$g_{PQ}$ & fraction\\ 
\hline
(a) & $\bino$ & $\widetilde{e}_1$ & 0.564 & 0.211 & 0.409 & 1.189 & 0.255 & 2.22 & 10.5 & 0.30 & 0.31 
%
\\
(b) &$\bino$ & $\widetilde{\nu}_e$ & 0.243 & 1.205 & 0.678 & 0.255 & 0.243 & 2.20 & 12.9 & 0.30 & 0.30  
%
\\
(c) & $\widetilde{\nu}_e$ & $\widetilde{e}_1$ & 0.440 & 0.673 & 0.439 & 1.004 & 1.201 & 0.31 & 13.0 & 0.11 & 0.080
%
\\
(d) & $\bino$ & $\widetilde{\tau}_1$ & 0.592 & 0.812 & 0.557 & 0.592 & 0.279 & 2.20 & 31.2 & 0.30 & 0.069 
%
\\
(e) &  $\widetilde{\tau}_1$ & $\bino$ & 0.448 & 0.401 & 0.549 & 0.370 & 0649 & 1.20 & 22.2 & 0.22 & 0.060
%
\\
(f) & $\bino$ & $\widetilde{t}_1$ & 0.205 & 0.573 & 0.851 & 0.717 & 0.550 & 0.45 & 11.4 & 0.13 & 0.037 
%
\\
(g) & $\bino$ & $\widetilde{\nu}_\tau$ & 0.697 & 0.237 & 0.483 & 0.999 & 0.590 & 2.37 & 31.1 & 0.31 & 0.035 
%
\\
(h) & $\widetilde{e}_1$ & $\bino$  & 0.846 & 0.252 & 0.470 & 0.988 & 0.507 & 2.40 & 18.1 & 0.31 & 0.026 
%
\\
(i) & $\widetilde{t}_1$ & $\bino$ & 0.800 & 0.520 & 0.500 & 0.931 & 0.933 & 2.21 & 19.0 & 0.30 & 0.019
%
\\
(j) & $\widetilde{\tau}_1$ & $\widetilde{e}_1$ & 0.562 & 0.369 & 0.675 & 0.283 & 0.457 & 1.13 & 11.7 & 0.21 & 0.018 
%
\\
(k) & $\widetilde{\nu}_\tau$ & $\widetilde{\tau}_1$ & 0.587 & 0.605 & 0.343 & 1.327 & 0.212 & 2.22 & 14.5 & 0.30 & 0.013 
%
\\
(l) & $\bino$ & $\widetilde{\chi}_1^\pm $ & 0.949 & 0.557 & 0.396 & 0.462 & 0.602 & 1.31 & 13.1 & 0.23 & 0.010 
%
\\
\hline
%
%
%
%
%
%
%
\end{tabular}
\end{center}
\hspace{10mm}\parbox{16.5cm}{\caption{Examples of $\la_i$,  $\widetilde{\D}$ and $\tan\beta$ leading to different NLSP/NNLSP patterns for $\xi=10^5\,\mathrm{GeV}$ and $F_X=10^{19}\,\mathrm{GeV}^2$. The lightest neutralino,
$\bino$ is always an almost  pure bino, except for the last example in which both the NLSP neutralino and the NNLSP chargino are higgsino-like. 
 }\label{t:ex}}
\end{table}
\section{Discussion}
\label{sec:disc}

We would like to start by enumerating similarities and differences between the
supersymmetric mass spectra obtained  here and
usually discussed models of gauge mediation. As a benchmark, we shall use
the characteristics of the supersymmetric mass spectra presented in \cite{primer}.

Similarities include: (i) gluinos are much heavier  than the lighter neutralinos, charginos and sleptons of the first two generations, (ii) thanks to large RGE effects squarks of the first two families are heavy as gluinos, (iii) the lighter stop is most often the lightest of squarks. 

We also find a number of notable differences from the usually considered spectra of GMSB.
%
%
%
%
%
%
%
%
%
%
For many cosmological and phenomenological purposes,
these differences 
can be most easily described by comparing the patterns of the NLSP/NNLSP particles.
In ordinary models of GMSB with the assumed scales $\xi$ and $M_Y$ and moderate (large) value of $\tan\beta$, one obtains bino/stau (stau/bino) patterns.
In our sample, bino/stau and stau/bino  are also frequent patterns (13\% of all cases),
but the lighter stau is often (approximately in 1 out of 3 cases) the superpartner of a left-handed state, such as in the example (d) in Table~\ref{t:ex}.

Our analysis reveals novel patterns with stop/sneutrino/selectron/higgsino (N)NLSP. 
%
%
Some of them 
may have advantages from the cosmological point of view.
A colored NLSP, such as the stop \cite{DiazCruz:2007fc} ($\sim$2\% in our sample), could help explain the problematic primordial abundance of lithium \cite{Kohri:2008cf}.
Although it may seem that the
stop mass range required by the lithium data leads to a larger stop relic abundance
than what is allowed by the Big Bang Nucleosynthesis \cite{Kusakabe:2010cc},
there may be additional
annihilations of stops after the QCD phase transitions when the stops are in a confined phase with quarks \cite{Kang:2006yd,Berger:2008ti}. The existing LHC data do not exclude  a light and long-lived stop
\cite{Kats:2011it}.

A bino NLSP, which has a large hadronic branching fraction in its decay, is also severely constrained by BBN: short lifetimes are preferred, so the gravitino mass cannot be too large, which in turn makes somewhat difficult to obtain a large reheating temperature \cite{Covi:2009bk}, unless the bino is degenerate in mass with a strongly interacting particle and the
two can coannihilate \cite{Covi:2010au} (see also \cite{Freitas:2007sa}). In our analysis we find a sizable fraction ($\sim$4\%) of models with the $\widetilde{B}/\widetilde{t}$
pattern; some of them exhibit a desired mass degeneracy. A parameter choice leading to
an appropriately degenerate spectrum is given as example (f) in Table \ref{t:ex}. For such a spectrum, we find with the use of the {\tt micrOmegas} code \cite{Belanger:2006is,Belanger:2008sj} a leading-order\footnote{This 
estimate neglects the Sommerfeld enhancement of the annihilation cross-section, which can give
a suppression of the relic density by a factor of $2-3$, cf.\ \cite{Covi:2010au}.} NLSP density parameter after freeze-out $\Omega_\mathrm{NLSP}\approx 10^{-2}$ , which
allows a reheating temperature consistent with the BBN bounds as large as $ 5\times 10^7\,\mathrm{GeV}$. This value is by 3 orders of magnitude larger than one obtained in models with a non-degenerate binos and a universal gaugino mass pattern at the high scale \cite{Covi:2010au}. 

The tension between BBN
and a large reheating temperature can be alleviated for a light sneutrino NLSP \cite{Kanzaki:2007pd} (see also \cite{Olechowski:2009bd}) -- such models are also present in our analysis, as shown by example (k) in Table \ref{t:ex}. In that example we find $\Omega_\mathrm{NLSP}=0.03$ \cite{Belanger:2006is,Belanger:2008sj} and the hadronic branching ratio in the sneutrino NLSP decay of $4\times 10^{-4}$, and conclude that the cosmological bounds
(BBN, CMB) diffuse neutrino background and are satisfied for the gravitino mass range
considered here \cite{Kanzaki:2007pd}.

Although the supersymmetric spectra described above are different from usually considered, they are compatible with the current LHC results  (see e.g.\ \cite{Sekmen:2011cz}), 
and we do not expect them to give any novel striking signal in the nearest future (with a possible exception of a charged detector-stable NLSP, a selectron or stau). Hopefully, with future data it will be possible to determine whether these
spectra can be realized in nature.

\section{Conclusions}
\label{sec:conc}

In this letter, we reported on a study of novel MSSM mass spectra predicted in a particular
unification model inspired by F-theory. In addition to bino/stau and stau/bino NLSP/NNLSP
pattern found in minimal GMSB models, we obtained a number of other patterns rarely discussed in the
literature. The degenerate bino/stop pattern and the pattern with a light NLSP sneutrino may have
some advantages from the cosmological point of view, as they allow for a much higher reheating temperatures than the usually considered models. In spite of the simplicity of the theoretical setup employed here, we obtain a great variety of possible supersymmetric spectra, which suggests that
the benchmark models used so far in the study of the LHC data may leave out some
realistic, well-motivated and cosmologically viable possibilities.

\subsubsection*{Acknowledgments}
\noindent \normalsize The authors would like to acknowledge stimulating discussions with M.~Badziak and S.~Lavignac 
and the use of a numerical code for calculating the 4-body sneutrino decay width developed
by S.~Trojanowski. We are especially indebted to P.~Chankowski for important comments and suggestions at the early stage of this paper. 
This work was partially supported by the MNiSW grant N N202 091839.

\newpage

\begin{thebibliography}{99}

\bibitem{Atlas-11-086}
ATLAS Collaboration, ATLAS-CONF-2011-086.

\bibitem{Sekmen:2011cz}
  S.~Sekmen {\it et al.},
  ``Interpreting LHC SUSY searches in the phenomenological MSSM,''
  arXiv:1109.5119 [hep-ph].

\bibitem{F-unify1}
 R. Donagi, M. Wijnholt, ``Model Building with F-Theory,'' [{arXiv:0802.2969 [hep-th]}];


\bibitem{F-unify2}
  C. Beasley, J.J. Heckman, C. Vafa, ``GUTs and Exceptional Branes in F-theory - I,''
   JHEP 0901:058 (2009), [{arXiv:0802.3391 [hep-th]}];
 
\bibitem{F-unify3}
   H. Hayashi et al.,
  ``New Aspects of Heterotic--F Theory Duality,''
  Nucl. Phys.  B 806:224 (2009), [{arXiv:0805.1057 [hep-th]}];
  
  \bibitem{F-unify4}
 C. Beasley , J.J. Heckman ,  C. Vafa,
  ``GUTs and Exceptional Branes in F-theory - II: Experimental Predictions,''
  JHEP 0901:059 (2009), [{arXiv:0806.0102 [hep-th]}];
  
\bibitem{Heckman:2009mn}
  J.~J.~Heckman, A.~Tavanfar, C.~Vafa,
  ``The Point of E(8) in F-theory GUTs,''
  JHEP {\bf 1008 } (2010)  040.
  [arXiv:0906.0581 [hep-th]].

\bibitem{Pawelczyk:2010xh}
  J.~Pawe{\l}czyk,
  ``F-theory inspired GUTs with extra charged matter,''
  Phys.\ Lett.\  {\bf B697 } (2011)  75-79.
  [arXiv:1008.2254 [hep-ph]].

\bibitem{Giudice:1998bp}
  G.~F.~Giudice, R.~Rattazzi,
  ``Theories with gauge mediated supersymmetry breaking,''
  Phys.\ Rept.\  {\bf 322 } (1999)  419-499.
  [hep-ph/9801271].  

\bibitem{Heckman:2010xz}
  J.~J.~Heckman, J.~Shao, C.~Vafa,
  ``F-theory and the LHC: Stau Search,''
  JHEP {\bf 1009 } (2010)  020.
  [arXiv:1001.4084 [hep-ph]].

 \bibitem{Dolan:2011aq}
  M.~J.~Dolan, J.~Marsano and S.~Schafer-Nameki,
  ``Unification and Phenomenology of F-Theory GUTs with $U(1)_{PQ},$''
  [arXiv:1109.4958 [hep-ph]].

  \bibitem{Dimopoulos:1996ig}
  S.~Dimopoulos and G.~F.~Giudice,
  ``Multi-messenger theories of gauge-mediated supersymmetry breaking,''
  Phys.\ Lett.\  B {\bf 393} (1997) 72
  [arXiv:hep-ph/9609344].

 \cite{Brahm:1990xx}
 \bibitem{Brahm:1990xx}
  D.~E.~Brahm, L.~J.~Hall, S.~D.~H.~Hsu,
  ``Ruling Out Large Sneutrino Vevs,''
  Phys.\ Rev.\  {\bf D42 } (1990)  1860-1862.

\bibitem{Ibe:2007km}
  M.~Ibe, R.~Kitano,
  ``Sweet Spot Supersymmetry,''
  JHEP {\bf 0708 } (2007)  016.
  [arXiv:0705.3686 [hep-ph]].

\bibitem{Chacko:2001km}
  Z.~Chacko, E.~Ponton,
  ``Yukawa deflected gauge mediation,''
  Phys.\ Rev.\  {\bf D66 } (2002)  095004.
  [hep-ph/0112190].

\bibitem{Giudice:1997ni}
  G.~F.~Giudice, R.~Rattazzi,
  ``Extracting supersymmetry breaking effects from wave function renormalization,''
  Nucl.\ Phys.\  {\bf B511 } (1998)  25-44.
  [hep-ph/9706540].
    
\bibitem{Shadmi:2011hs}
  Y.~Shadmi, P.~Z.~Szabo,
  ``Flavored Gauge-Mediation,''
  [arXiv:1103.0292 [hep-ph]].
  
  \bibitem{Evans:2011be}
  J.~L.~Evans, M.~Ibe, T.~T.~Yanagida,
  ``Relatively Heavy Higgs Boson in More Generic Gauge Mediation,'' 
  [arXiv:1107.3006 [hep-ph]].
  
\bibitem{Dine:1996xk}
  M.~Dine, Y.~Nir, Y.~Shirman,
  ``Variations on minimal gauge mediated supersymmetry breaking,''
  Phys.\ Rev.\  {\bf D55 } (1997)  1501-1508.
  [hep-ph/9607397].  
  
\bibitem{Djouadi:2002ze}
  A.~Djouadi, J.~-L.~Kneur, G.~Moultaka,
  ``SuSpect: A Fortran code for the supersymmetric and Higgs particle spectrum in the MSSM,''
  Comput.\ Phys.\ Commun.\  {\bf 176 } (2007)  426-455.
  [hep-ph/0211331].    
  
\bibitem{Kitano:2006wz}
  R.~Kitano,
  ``Gravitational Gauge Mediation,''
  Phys.\ Lett.\  {\bf B641 } (2006)  203-207.
  [hep-ph/0607090].  

\bibitem{LeMouel:2001sf}
  C.~Le Mouel,
  ``Optimal charge and color breaking conditions in the MSSM,''
  Nucl.\ Phys.\  {\bf B607 } (2001)  38-76.
  [hep-ph/0101351].

\bibitem{primer}
  S.~P.~Martin,
  ``A Supersymmetry primer,''
  In *Kane, G.L. (ed.): Perspectives on supersymmetry* 1-98.
  [arXiv:hep-ph/9709356 [hep-ph]].

  \bibitem{DiazCruz:2007fc}
  J.~L.~Diaz-Cruz, J.~R.~Ellis, K.~A.~Olive and Y.~Santoso,
  ``On the Feasibility of a Stop NLSP in Gravitino Dark Matter Scenarios,''
  JHEP {\bf 0705} (2007) 003
  [arXiv:hep-ph/0701229].

\bibitem{Kohri:2008cf}
  K.~Kohri, Y.~Santoso,
  ``Cosmological scenario of stop NLSP with gravitino LSP and the cosmic lithium problem,''
  Phys.\ Rev.\  {\bf D79 } (2009)  043514.
  [arXiv:0811.1119 [hep-ph]].

  \bibitem{Kusakabe:2010cc}
  M.~Kusakabe, T.~Kajino, T.~Yoshida and G.~J.~Mathews,
  ``Big Bang Nucleosynthesis with long-lived strongly interacting relic
  particles,''
  arXiv:1001.1413 [astro-ph.CO].

  \bibitem{Kang:2006yd}
  J.~Kang, M.~A.~Luty and S.~Nasri,
  ``The Relic abundance of long-lived heavy colored particles,''
  JHEP {\bf 0809} (2008) 086
  [arXiv:hep-ph/0611322].
  
  \bibitem{Berger:2008ti}
  C.~F.~Berger, L.~Covi, S.~Kraml and F.~Palorini,
  ``The Number density of a charged relic,''
  JCAP {\bf 0810} (2008) 005
  [arXiv:0807.0211 [hep-ph]].

\bibitem{Kats:2011it}
  Y.~Kats, D.~Shih,
  ``Light Stop NLSPs at the Tevatron and LHC,''
  JHEP {\bf 1108 } (2011)  049.
  [arXiv:1106.0030 [hep-ph]].

\bibitem{Covi:2009bk}
  L.~Covi, J.~Hasenkamp, S.~Pokorski and J.~Roberts,
  ``Gravitino Dark Matter and general neutralino NLSP,''
  JHEP {\bf 0911} (2009) 003
  [arXiv:0908.3399 [hep-ph]].

 \bibitem{Covi:2010au}
  L.~Covi, M.~Olechowski, S.~Pokorski, K.~Turzynski and J.~D.~Wells,
  ``Supersymmetric mass spectra for gravitino dark matter with a high reheating
  temperature,''
  JHEP {\bf 1101} (2011) 033
  [arXiv:1009.3801 [hep-ph]].

  \bibitem{Freitas:2007sa}
  A.~Freitas,
  ``Radiative corrections to co-annihilation processes,''
  Phys.\ Lett.\  B {\bf 652} (2007) 280
  [arXiv:0705.4027 [hep-ph]].

  \bibitem{Belanger:2006is}
  G.~Belanger, F.~Boudjema, A.~Pukhov and A.~Semenov,
  ``MicrOMEGAs 2.0: A Program to calculate the relic density of dark matter in
  a generic model,''
  Comput.\ Phys.\ Commun.\  {\bf 176} (2007) 367
  [arXiv:hep-ph/0607059].
  
  \bibitem{Belanger:2008sj}
  G.~Belanger, F.~Boudjema, A.~Pukhov, A.~Semenov,
  ``Dark matter direct detection rate in a generic model with micrOMEGAs 2.2,''
  Comput.\ Phys.\ Commun.\  {\bf 180 } (2009)  747-767.
  [arXiv:0803.2360 [hep-ph]].

  \bibitem{Kanzaki:2007pd}
  T.~Kanzaki, M.~Kawasaki, K.~Kohri and T.~Moroi,
  ``Cosmological Constraints on Neutrino Injection,''
  Phys.\ Rev.\  D {\bf 76}, 105017 (2007)
  [arXiv:0705.1200 [hep-ph]].

                  
  \bibitem{Covi:2007xj}
  L.~Covi and S.~Kraml,
  ``Collider signatures of gravitino dark matter with a sneutrino NLSP,''
  JHEP {\bf 0708} (2007) 015
  [arXiv:hep-ph/0703130].

\bibitem{Olechowski:2009bd}
  M.~Olechowski, S.~Pokorski, K.~Turzynski and J.~D.~Wells,
  ``Reheating Temperature and Gauge Mediation Models of Supersymmetry
  Breaking,''
  JHEP {\bf 0912} (2009) 026
  [arXiv:0908.2502 [hep-ph]].


  

  



  


  

     

  
  
\end{thebibliography}
\end{document}